\documentclass[aps,preprint,groupedaddress,superscriptaddress]{revtex4}

\usepackage{graphicx}
\usepackage{dcolumn}
\usepackage{bm}
\usepackage{color}

\begin{document}



\title{Exploring interlayer coupling in the twisted bilayer PtTe$_{2}$ }
\author{Jeonghwan Ahn$^{\dag}$}
\affiliation{Materials Science and Technology Division, Oak Ridge National Laboratory, Oak Ridge, Tennessee 37831, USA}
\thanks{These authors contributed equally to this work.}
\author{Seoung-Hun Kang$^{\dag}$}
\affiliation{Materials Science and Technology Division, Oak Ridge National Laboratory, Oak Ridge, Tennessee 37831, USA}
\thanks{These authors contributed equally to this work.}
\author{Mina Yoon}
\affiliation{Materials Science and Technology Division, Oak Ridge National Laboratory, Oak Ridge, Tennessee 37831, USA}
\author{Jaron T. Krogel}
\email{krogeljt@ornl.gov}
\affiliation{Materials Science and Technology Division, Oak Ridge National Laboratory, Oak Ridge, Tennessee 37831, USA}

\date{\today}


\begin{abstract}
{
We have investigated interlayer interactions in the bilayer PtTe$_{2}$  system, which influence the electronic energy bands near the Fermi levels.
Our diffusion Monte Carlo (DMC) calculations for the high-symmetry bilayer stackings (AA, AB, AC) manifest distinct interlayer binding characteristics among the stacking modes by revealing significantly different interlayer separations depending on the stackings, which is critical to understanding the interlayer coupling of the twisted bilayers consisting of various local stacking arrangements.
Furthermore, a comparison between the interlayer separations obtained from DMC and density functional theory (DFT) shows that meta-GGA-based vdW-DFT results agree with DMC for different layer stackings, including twisted bilayers, but only the ground-state AA stacking matches well with GGA-based DFT predictions. 
This underscores the importance of accurate exchange-correlation potentials even for capturing the stacking-dependent interlayer binding properties.
We further show that the variability in DFT-predicted interlayer separations is responsible for the large discrepancy of band structures in the 21.79$^{\circ}$-twisted bilayer PtTe$_{2}$ , affecting its classification as metallic or semiconducting. These results demonstrate the importance of obtaining a correct description of stacking-dependent interlayer coupling in modeling delicate bilayer systems at finite twists. 
}


\end{abstract}


\maketitle


\section{Introduction}
\label{sec:Introduction}

The nature of interlayer coupling is pivotal for two-dimensional (2D) layered materials since layer stacking has unveiled emergent features never seen in their single-layer form. Several stacking fashions when layering the 2D materials allow us to explore their novel properties by manipulating stacking configurations away from the natural ground-state~\cite{cao2018correlated,vizner2021interfacial,hagymasi2022observation}, which leads to weakened interlayer coupling within the system. Subtle metastability arising from altering interlayer coupling enriches the correlated electronic phases in 2D materials, as exemplified by different topological properties depending on the stacking fashions~\cite{qian2014quantum,soluyanov2015type,chang2016prediction,kim2017origins,kang2022reshaped}. These characteristics underscore the key role of interlayer coupling in understanding the properties of 2D layered materials.

Among the various layered materials, PtTe$_{2}$ has attracted a great deal of attention due to the presence of the strong interlayer coupling resulting in its intriguing electronic properties, such as topological semimetals~\cite{Zhang17,yan2017lorentz} and the highest electrical conductivity at room temperature among metallic transition metal dichalcogenides (TMDs)~\cite{hao2018low,xu2020high}. 
Interestingly, a semiconducting single-layer PtTe$_{2}$ undergoes a transition to a semimetallic bilayer structure and maintains its semimetallicity up to the bulk structure~\cite{lin2020dimensionality} as a consequence of the interlayer hybridization between Te-p$_{z}$ orbitals leading to significant band gap reduction when forming the layered structures.
This highlights the importance of the interactions between layers in PtTe$_{2}$.
Furthermore, the interlayer coupling is predicted to be controlled by a lateral shift of the upper layer with respect to the bottom layer, prospecting its potential uses in technologies like spintronics~\cite{li2021tunable} and ferroelectrics~\cite{sheng2023ferroelectric}.

More recently, the interlayer coupling of twisted systems in 2D layered materials has been studied intensively, as the twist gives rise to an additional degree of freedom to promote correlated phenomena of the system.
The twisted layers are structured by the various stacking domains ranging from the most stable to least stable arrangements, with each local domain having different interlayer coupling effects. The presence of low stability local stackings in the twisted system results in weakening the interlayer interactions overall as the interlayer separation increases relative to the untwisted ground state stacking. The degree of interlayer coupling thus evolves as a function of the twist angle~\cite{liu2014evolution,yan2019probing,zheng2023strong} and the tunable interlayer coupling in the twisted bilayer systems opens a new possibility of manipulating the properties of materials by twist angles such as exciton dynamics dependent on twist angles in MoSe$_{2}$/WSe$_{2}$ heterostructures~\cite{nayak2017probing}.
Moreover, for PtTe$_{2}$, the way the layers interact is considered a mix of different forces originating from not only van der Waals (vdW) interactions but also the interlayer hybridization between the tellurium atoms in neighboring layers in addition to the possibility for metallic bonding. The complexity of the interactions alongside the structural features of twisted bilayer PtTe$_{2}$  makes it challenging for theory to accurately describe interlayer interactions in bilayer PtTe$_{2}$ . 

The challenges in modeling interlayer interactions are manifested in density functional theory (DFT) calculations showing a notable difference in the interlayer distances for PtTe$_{2}$~\cite{li2021tunable}, and its cousin materials of PtS$_{2}$~\cite{zhao2016extraordinarily,ahmad2020strain} and PtSe$_{2}$~\cite{li2021layer,zhang2021precise}.
As the interlayer distance is understood as a crucial parameter to directly tune the interlayer coupling~\cite{liu2014evolution,nayak2017probing}, this variation, alongside the lack of experimental measurements for the bilayer structure of PtTe$_{2}$ makes choosing reliable DFT exchange-correlation functionals difficult.
The many-body diffusion Monte Carlo (DMC) method, which can describe all types of bonding equally by explicitly incorporating the interactions between electrons, has been shown to overcome the challenges in describing complicated interlayer interactions.
DMC has been utilized as a reliable reference, especially when there is no experimental data available for 2D layered materials~\cite{mostaani15,shulenburger15,shin17,ahn18,ahn21b,ahn2023structural}. This has led us to employ DMC to assist in choosing the most reliable exchange-correlation functionals within DFT to further enable broader studies of twisted PtTe$_{2}$ systems. 

In this work, we investigate interlayer interactions in bilayer PtTe$_2$ using both DFT and DMC calculations. 
A comparison between the interlayer distances obtained from DMC and various DFT calculations shows that the r$^{2}$SCAN-based vdW functional results match DMC closely for different layer arrangements, including twisted bilayers. GGA-based calculations, however, only match well for one specific arrangement (the ground state stacking) and show a wide range of layer distances for others with direct impacts on the predicted properties of the system.
This inconsistency highlights the necessity to accurately model how layers interact across a variety of different stacking arrangements. This need is acute for twisted bilayers, which consist of various local stacking arrangements where the interlayer interactions differ significantly.

\section{Methodology}
\label{sec:Methodology}
We have performed the DFT calculations using the VASP package~\cite{Kresse1993,Kresse1996}. The projector-augmented wave pseudopotentials~\cite{PAW1994,Kresse1999} for Pt and Te atoms were used to solve the Kohn-Sham equations by incorporating spin-orbit coupling effects. A plane-wave cutoff was set to be 400 eV, and a primitive Brillouin zone was sampled with $12 \times 12 \times 1$ ($12 \times 12 \times 8$) Monkhorst–Pack~\cite{monkhorst76} $k$ mesh for the bilayer and monolayer (bulk). The structural optimizations were done with a force threshold of $5 \times 10^{-3}$ eV/\AA~ while electronic convergence was achieved with a threshold of 10$^{-6}$ eV. To avoid spurious interactions between periodic images, a vacuum distance for the monolayer and bilayer is set to be 30~\AA~ along the perpendicular direction to the PtTe$_{2}$ plane. 
In this study, we consider self-consistent vdW functionals with different combinations of GGA exchange and the non-local correlation functionals such as vdW-optB86b~\cite{vdW-optB88}, vdW-optB88~\cite{vdW-optB86b}, vdW-DF2~\cite{vdW-DF2}, rev-vdW-DF2~\cite{rev-vdW-DF2}, rVV10~\cite{rVV10} as well as dispersion-corrected schemes including PBE+D3~\cite{grimme10}, and PBE+MBD~\cite{MBD1}. Furthermore, vdW functionals in the combination of the meta-GGA functional which additionally incorporates the local kinetic energy density term into the exchange part such as r$^{2}$SCAN+rVV10~\cite{r2SCAN_rVV10}, r$^{2}$SCAN+D3 and r$^{2}$SCAN+MBD are considered, which allows us to discuss importance of the semi-local part in the vdW-DFT calculations.

Our fixed-node DMC calculations were performed with the QMCPACK package~\cite{kim18,kent20}. We used a trial wave function of the Slater-Jastrow form~\cite{jastrow55} to incorporate many-body correlation effects into the Jastrow factors. In this study, we considered electron-ion, electron-electron, and electron-electron-ion correlation effects described by one-, two- and three-body Jastrow factors, respectively. The Slater determinant comprised the PBE orbitals by solving the Kohn-Sham equations~\cite{kohn65} with the QUANTUM ESPRESSO~\cite{giannozzi09}. The corresponding DFT calculations were performed with a plane-wave cutoff of 400Ry and $18 \times 18 \times 1$ Monkhorst-Pack $k$-point mesh~\cite{monkhorst76} for the monolayer and bilayer structures using a norm-conserving potential with pseudovalence state of $5s^{2}5p^{6}6s^{1}5d^{9}$ for Pt atom, and the one developed within the ccECP scheme~\cite{bennett2017new,annaberdiyev2018new} for Te atom, which were utilized in our previous DMC studies involved with these atomic species~\cite{ahn20,bennett2022magnetic,ahn2023structural,ahn2023stacking}. 
The one-body size effects were eliminated through the twist-averaged boundary conditions~\cite{lin01}, and 36, 16, and 9 twist angles were utilized for three different supercell sizes of $2 \times 2 \times 1$, $3 \times 3 \times 1$, and $4 \times 4 \times 1$, respectively. The two-body size effects were mitigated by the extrapolation to the thermodynamic limit ($N = \infty$) by linear fit with the twisted-averaged total energies of three different supercells (See Fig.1 of Supplemental Material). The Jastrow parameters were optimized through variational Monte Carlo calculations based on the linear method proposed by Umrigar {\it et al.}~\cite{umrigar07} while subsequent DMC calculations were carried out using a conservative time step of $\tau = 0.005$ Ha$^{-1}$ in combination with size-consistent T-moves for variationally estimating non-local term of pseudopotentials as imaginary time projection proceeded~\cite{kim18}.

\section{Results}

\begin{figure*}[h]
\centering
\includegraphics[width=6.5in]{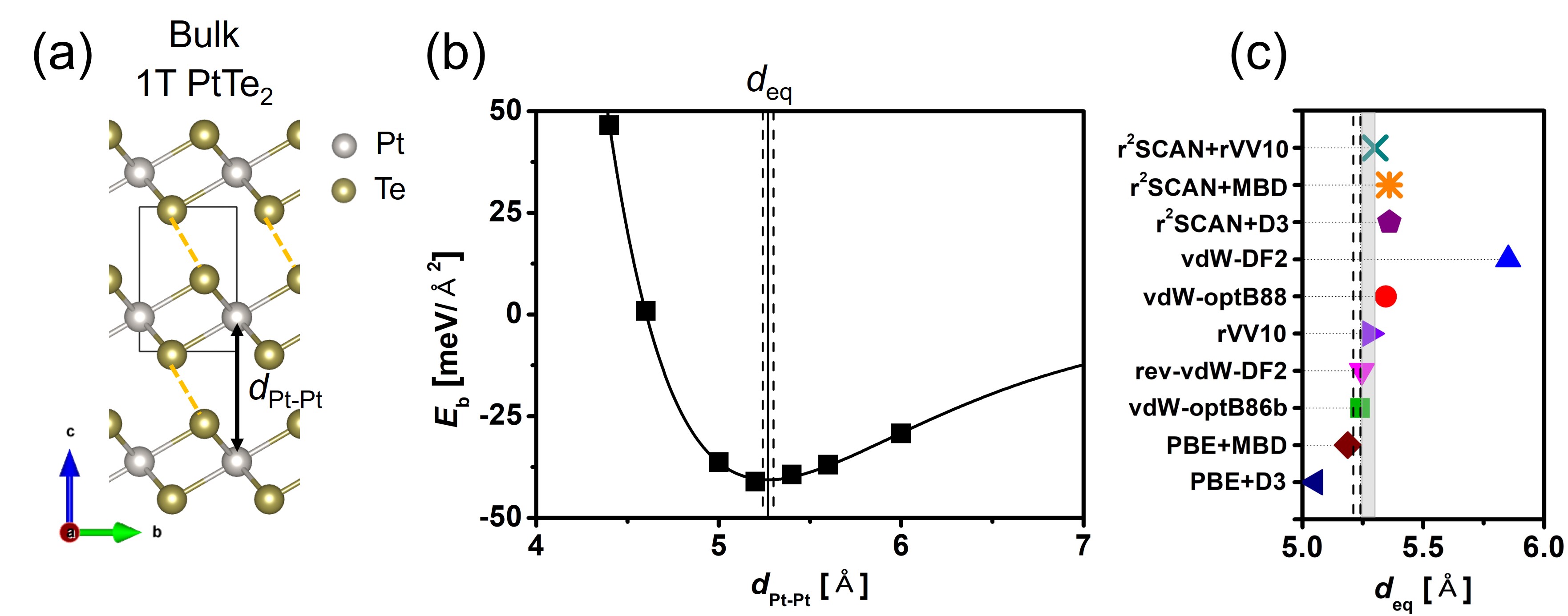}
\caption{(a) Side view of the crystal structure of bulk PtTe$_{2}$. The primitive cell is denoted by a black rectangle, and the corresponding interlayer separation is denoted by $d_{\text{Pt-Pt}}$. (b) DMC interlayer binding curve for bulk PtTe$_{2}$ computed with $3 \times 3 \times 2$ supercell. The equilibrium interlayer separation $d_{\text{eq}}$ is represented by a solid vertical line, while dotted vertical lines represent its statistical uncertainty. (c) Comparison between equilibrium interlayer separations from DMC, DFT, and experiment. The DMC data in (c) is represented by a gray-colored region reflecting the statistical uncertainty, while the experimental values are denoted by vertical dotted lines. 
}
\label{fig:bulk}
\end{figure*}

\subsection{DMC benchmark of bulk PtTe$2$}
Figure~\ref{fig:bulk} (a) shows that the crystal structure of bulk PtTe$_{2}$ consisting of a single layer of PtTe$_2$ structured in three atomic layers. This structure features a TMD, each transition metal layer sandwiched between two chalcogen atom layers, forming an octahedral unit of 1T structure. This layered material has an AA stacking configuration in its bulk structure, confirmed through experimental observations~\cite{Zhang17,lin2020dimensionality}. 
Figure~\ref{fig:bulk}(c) presents interlayer separations computed with several different DFT exchange-correlation functionals in comparison with experiment~\cite{Zhang17,lin2020dimensionality} for the bulk PtTe$_{2}$. From this, we first observed that several DFT functionals like r$^2$SCAN+rVV10, r$^2$SCAN+D3, r$^2$SCAN+MBD, vdW-optB86b, rev-vdW-DF2, rVV10, and PBE+MBD produced results within the range of the experiments. In contrast, the PBE+D3 and vdW-DF2 calculations showed larger deviations from these values.
Next, we examine whether DMC can produce reasonable results comparable to bulk experimental values by computing the interlayer binding curve for bulk PtTe$_{2}$ as shown in Fig.~\ref{fig:bulk}(b). For this DMC calculation, we utilized a $3 \times 3 \times 2$ supercell containing 540 electrons where the in-plane lattice constant, $a$, is set to be the experimental value of 4.01~\AA~\cite{lin2020dimensionality}. The equilibrium DMC interlayer distance is 5.27(3)~\AA, corresponding to the distance yielding maximum interlayer binding energy, which closely aligns with the experimentally observed distance of approximately 5.20~\AA~\cite{kliche1985far} to 5.24~\AA~\cite{lin2020dimensionality}, as shown in Fig.~\ref{fig:bulk}(c). This demonstrates the reliability of the DMC method for studying interlayer interactions in layered PtTe$_2$ materials, which justifies its role of acting as a reference relative to DFT for the bilayer systems that lack experimental measurement for the structural parameters.

\begin{figure*}
\centering
\includegraphics[width=6.5in]{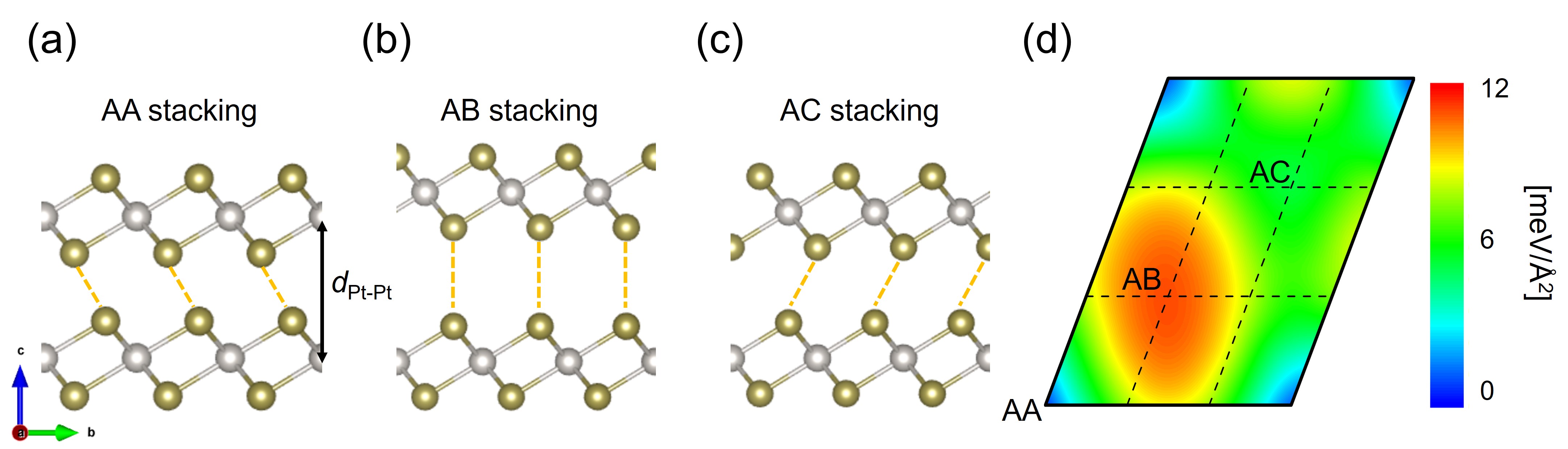}
\caption{Side view of the crystal structure for the (a) AA-, (b) AB-, and (c) AC-stacked bilayer PtTe$_{2}$ s. The corresponding interlayer separation $d_{\text{Pt-Pt}}$ is denoted in (a). (d) Potential energy surface for the lateral translations of the upper PtTe$_{2}$ layer with respect to the lower one computed with r$^{2}$SCAN+rVV10.}
\label{fig:DFT_energy}
\end{figure*}

\subsection{Discrepancies in DFT energetics and interlayer separations of high symmetry bilayer stackings}
We now analyze the bilayer PtTe$_{2}$  systems. One can consider three different high-symmetric stacking modes, namely, AA, AB, and AC stacking modes, as displayed in Fig.~\ref{fig:DFT_energy}(a), (b), and (c), respectively. These stacking modes can be identified in Fig.~\ref{fig:DFT_energy}(d), which presents a contour map of relative total energies computed with r$^{2}$SCAN+rVV10 for the lateral shift of the upper PtTe$_{2}$ layer with respect to the lower layer (the potential energy surface of relative lateral translations). We find that AA stacking is the lowest energy configuration, and AC stacking corresponds to a very shallow local minimum in the stacking energy maps (a metastable state), with the energy difference from the ground-state structure being 3.5 meV/\AA$^{2}$ In contrast, the AB stacking is the most unstable configuration.
Note that the overall energy ordering among the high symmetry stacking modes is not affected by the choice of the DFT functional.
Nevertheless, there are large quantitative inconsistencies in the energetics among different bilayer stacking modes, reflecting the different descriptions of the interlayer interactions arising from the metallic + vdW + hybridization type bonding in PtTe$_{2}$ depending on the DFT functional. Therefore, high-level benchmarking to the many-body DMC calculations is desirable to arrive at a trustworthy prediction of the stability of the bilayer structures with different stacking modes.

\begin{figure*}
\centering
\includegraphics[width=6.5in]{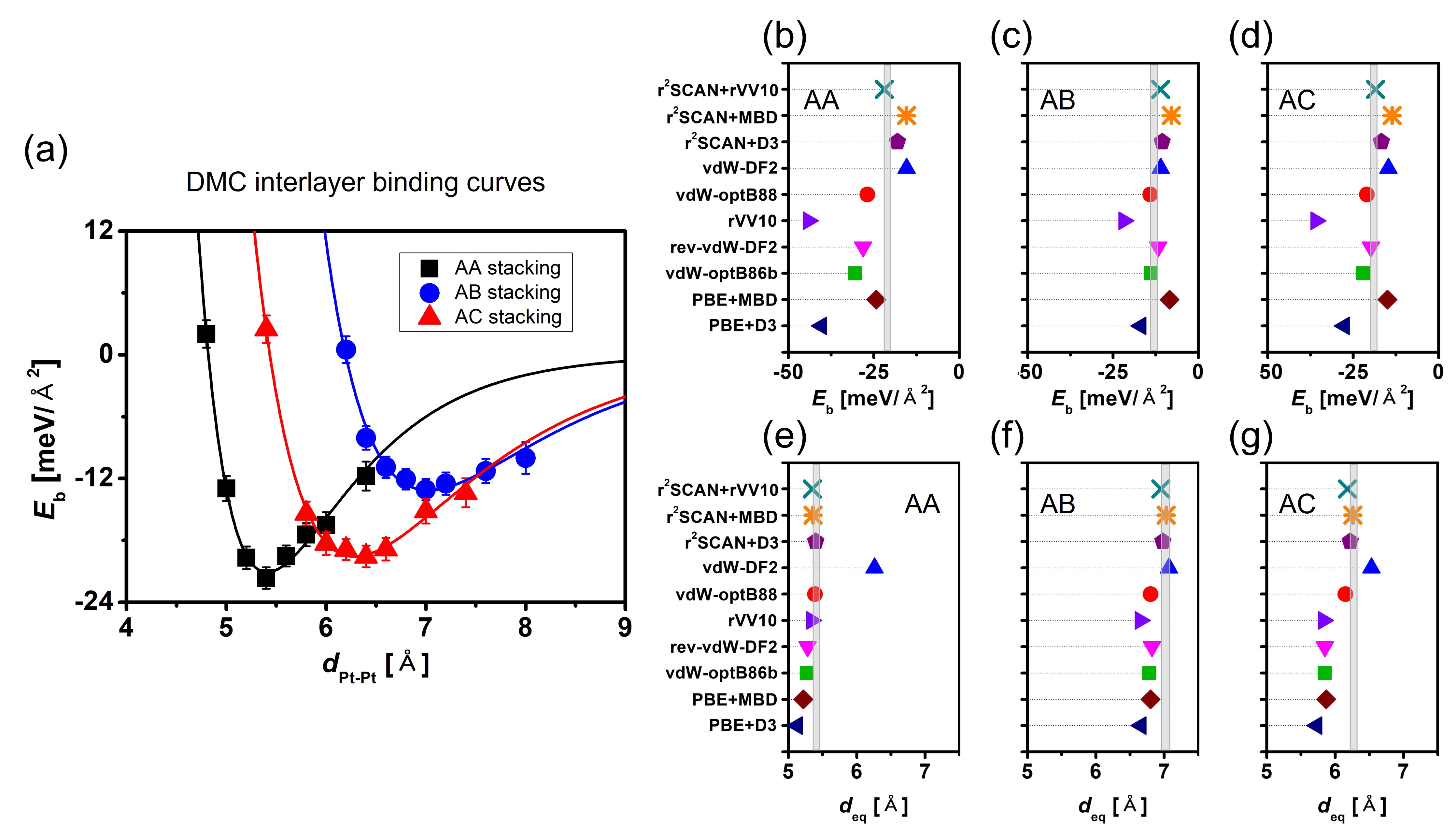}
\caption{(a) DMC interlayer binding curve for the AA-, AB-, and AC-stacked bilayer PtTe$_{2}$ s constructed with the finite-size corrected DMC energies. 
The comparison between DMC and DFT interlayer binding energies for (b) AA, (c) AB, and (d) AC stacking modes while the comparison of interlayer separations for AA-, AB-, and AC-stacked bilayer PtTe$_{2}$  are presented in (e), (f), and (g). 
The DMC data in (b)-(g) is represented by a gray-colored region reflecting the statistical uncertainty.}
\label{fig:QMC_data}
\end{figure*}

To this end, we performed the DMC calculations for AA-, AB-, and AC-stacked bilayer PtTe$_{2}$  as representative elements of stable, unstable, and intermediate stacking regions, respectively, to understand the interlayer coupling in the generic stacking modes as well as even twisted bilayer system. Figure~\ref{fig:QMC_data}(a) presents the DMC interlayer binding curves for three stacking modes, which were constructed with finite-size corrected DMC energies extrapolated to the thermodynamic limit as in Fig. 1 of Supplemental Material~\cite{SI}.
This allows us to determine equilibrium interlayer binding energies for each stacking mode, which confirms AA stacking to be the most stable among the three symmetric stacking modes, as predicted in our DFT calculations discussed above. The interlayer binding energies for AB and AC stacking modes are higher than that of the AA stacking by 8(1) and 2(1) meV/\AA$^{2}$, respectively. The DMC binding energies are compared with the DFT binding energies computed with several different vdW functionals for each stacking as presented in Fig.~\ref{fig:QMC_data}(b)-(d) and Table 1 of Supplemental Material~\cite{SI}. The comparison shows that the results from the r$^{2}$SCAN-based functionals are consistently superior to the GGA-based results, which is understood to be due to the improvement of inhomogeneity from incorporating the kinetic energy density into the meta-GGA functionals.
However, among them, r$^{2}$SCAN+rVV10 results best agree with DMC in predicting the binding energies for all stacking modes, which makes it robust even in predicting the relative energies among the bilayer structure with different stacking modes.

\begin{figure*}[h]
\centering
\includegraphics[width=6.5in]{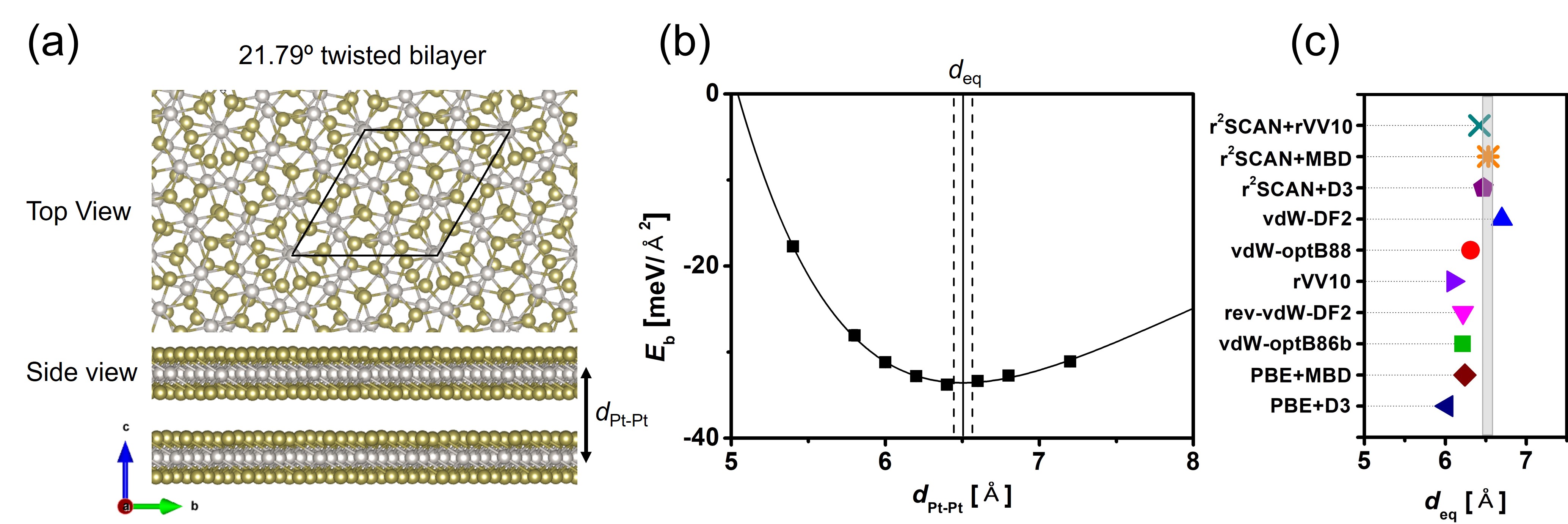}
\caption{(a) Top and side views of the crystal structure of 21.79$^{\circ}$ twisted bilayer PtTe$_{2}$ . The primitive unit cell is represented with the parallelogram enclosed by black lines, and the corresponding interlayer separation is denoted by $d_{\text{Pt-Pt}}$.
(b) DMC interlayer binding curve for 21.79$^{\circ}$ twisted bilayer PtTe$_{2}$  computed with $1 \times 1 \times 1$ supercell. The equilibrium interlayer separation $d_{\text{eq}}$ is represented by a solid vertical line, while dotted vertical lines represent its statistical uncertainty.
(c) Comparison of DMC and DFT interlayer separations for 21.79$^{\circ}$ twisted bilayer PtTe$_{2}$ . The DMC data in (c) is represented by a gray-colored region reflecting the statistical uncertainty.}
\label{fig:2179_binding_curve}
\end{figure*}


Next, let us consider interlayer separations, which characterize the nature of the interlayer coupling, for the symmetrically stacked and twisted bilayer PtTe$_{2}$  from the comparison between DMC and DFT results. The interlayer separation is necessary to be more carefully investigated than the interlayer binding energies because it is critically relevant to the electronic structures especially near the Fermi level, which makes it a crucial factor resulting in the sensitive electronic properties. Figure~\ref{fig:QMC_data}(e)-(g) displays interlayer separations computed with DMC and DFT for AA-, AB-, and AC-stacked bilayer PtTe$_{2}$  (the corresponding data is tabulated in Table 2 of Supplemental Material~\cite{SI}). The DMC interlayer binding distance for AA stacking mode is found to be the shortest, followed by those of AC and AB stacking modes, with 5.41(5)~\AA, 6.27(5)~\AA, and 7.02(6)~\AA, respectively. 
The considerable variation in the bilayer separation distances suggests the characteristics of the interlayer coupling would be significantly different depending on the stacking modes, manifesting the strong interlayer coupling in the bilayer PtTe$_{2}$  system. Indeed, the interlayer separations for the most stable AA and the most unstable AB stackings correspond to the interlayer nearest neighbor Te-Te distances of 3.50(5)~\AA~ and 4.24(6)~\AA~, respectively. These distances are within a range from the sum of the covalent radii of Te atoms (2.76~\AA)~\cite{cordero2008covalent} to the vicinity of the sum of their vdW radii (4.12~\AA)~\cite{bondi1964van}, suggesting that orbital hybridization effects are substantially involved in interlayer couplings in addition to the vdW interactions for the AA stacking, while the vdW interactions mainly characterize the interlayer coupling for the AB stacking. 
The interlayer hybridization of AA stacking is further evidenced by the agreement for the corresponding interlayer separations between DMC and semilocal functionals of PBE (5.36~\AA) and r$^2$SCAN (5.47~\AA), in which the long-range vdW correlation is absent.

As shown in Fig.~\ref{fig:2179_binding_curve}(e)-(g) presenting the comparison between DMC and DFT results, one can see that r$^{2}$SCAN+rVV10, r$^{2}$SCAN+MBD, r$^{2}$SCAN+D3, vdW-optB88, and rVV10 results agree with DMC for the AA stacked bilayer. However, for the AB and AC stacking modes, only r$^{2}$SCAN-based vdW functionals match with DMC, while vdW-optB88 and rVV10 separation distances become far away from the DMC values, including the rest DFT functionals that poorly describe the AA stacking.
This indicates that the stacking-dependent interlayer coupling is similarly described with the DMC and the r$^{2}$SCAN-based vdW functionals overall, possibly due to its improvement of the semi-local part in the vdW functionals, which highlights the importance of a role of semi-local exchange-correlation potential even in the prediction of the interlayer binding properties.
More specifically, the discrepancy in the DFT interlayer separations is found to be largest for the AC stackings, followed by the AB and AA stackings. Considering that the binding natures of AA and AB stackings turn out to be mainly characterized by interlayer Te-p$_{z}$ hybridization and the vdW interactions, DFT is considered to have more limitations in describing the interlayer interactions with both contributions being comparable to each other such as in the case of the AC stacking. This strongly suggests that the different degrees of discrepancy in the DFT interlayer separations depending on the stacking modes would still influence that of a twisted bilayer PtTe$_{2}$  in light of its structural feature consisting of various local stacking domains, including AA-, AB-, and AC-like regions.

\subsection{Discrepancies in DFT interlayer separations of twisted bilayer}

To further examine the performance of candidate DFT functionals for a twisted system, we investigate the twisted bilayer PtTe$_{2}$  with a twist angle of 21.79$^{\circ}$ from the AA stacking. It has the smallest Moir{\'e} pattern among the twisted bilayer systems, namely, p($\sqrt{7} \times \sqrt{7}$), as shown in Fig.~\ref{fig:2179_binding_curve}(a). When constructing the DMC interlayer binding curves as displayed in Fig.~\ref{fig:2179_binding_curve}(b) for the 21.79$^{\circ}$ twisted bilayer, we used the $1 \times 1 \times 1$ unit cell containing 420 electrons. Figure~\ref{fig:2179_binding_curve}(c) presents the optimized interlayer binding distance computed from DMC and DFT calculations. The DMC calculations find the equilibrium distances to be 6.52(6)~\AA, which falls between that of the AB and AC stackings. This is a consequence of the mixed stacking configurations from stable to unstable ones. In terms of interlayer separations, the 21.79$^{\circ}$ twisted bilayer is expected to show interlayer hybridization plus vdW characteristics that are slightly weaker than that of the AC stacking mode, underscoring the importance of the well-balanced description of interlayer hybridization and vdW interactions to simulate the twisted bilayer PtTe$_{2}$  accurately.

From the comparison between DMC and DFT results for the 21.79$^{\circ}$ twisted bilayer as shown in Fig.~\ref{fig:2179_binding_curve}(d), r$^{2}$SCAN-based vdW functionals that are robust for each high-symmetry stacking mode also agree with the DMC interlayer separation for the twisted bilayer PtTe$_{2}$ . 
However, in conjunction with the energetics discussed above, we conclude that r$^{2}$SCAN+rVV10 is the most reliable for simulating the PtTe$_{2}$ systems, even including twist, among the currently available vdW-corrected functionals.
Interestingly, the discrepancy in the DFT interlayer separations for the 21.79$^{\circ}$ twisted bilayer is similarly patterned to that of the AB and AC stacking modes, which have interlayer separations closer to that of the twisted system as compared to AA stacking. From this fact, we understand the large discrepancy for the twisted bilayer to originate from larger discrepancies for the AB and AC stackings and not from the equilibrium AA stacking mode. This highlights the importance of correctly describing the interlayer couplings varying with the stacking modes away from the ground-state AA stacking to include less stable configurations such as AB and AC stackings when it comes to understanding twisted bilayer systems.
The large discrepancy for the twisted bilayer can also be understood in terms of its optimal DMC interlayer separation of 6.52(6)~\AA, where the contributions from the interlayer hybridization and vdW interactions are expected to be comparable. However, as discussed above, it turns out to be challenging to describe with the DFT calculations.

\begin{figure*}
\centering
\includegraphics[width=6.5in]{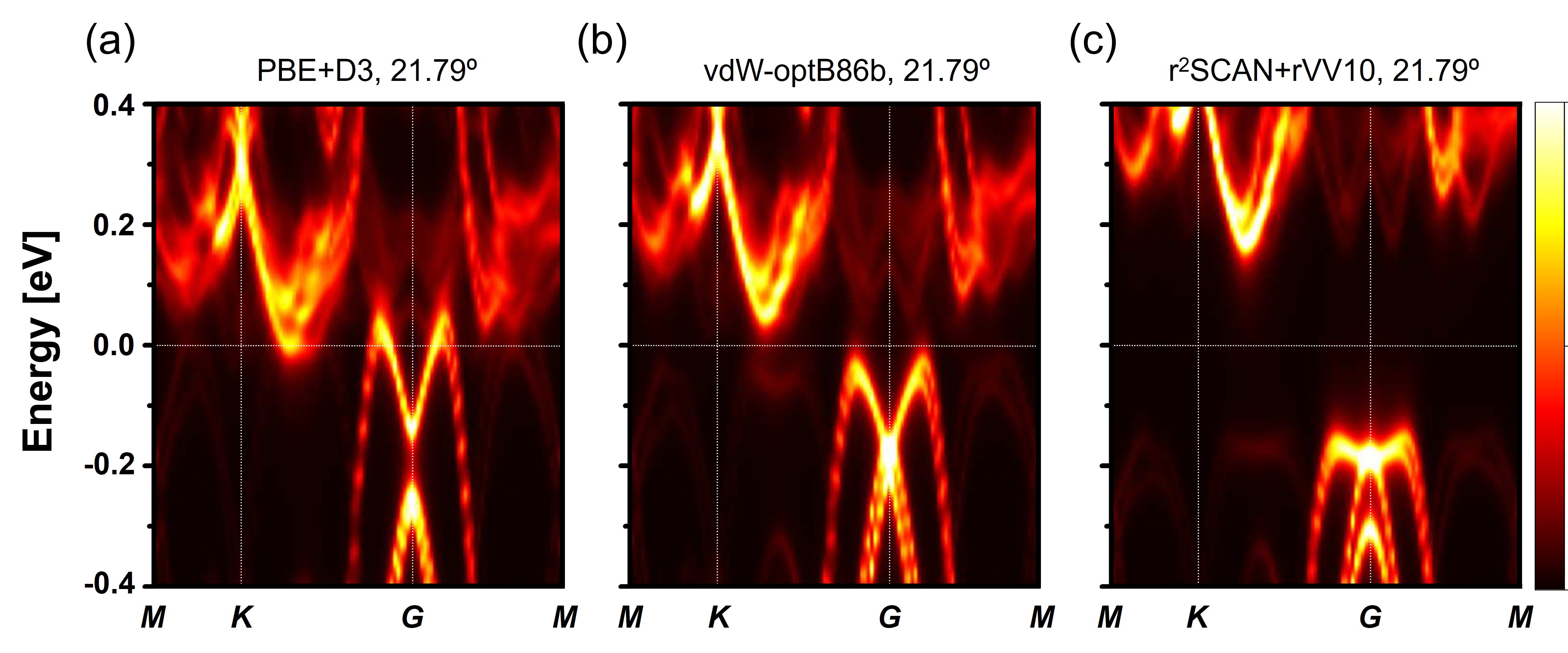}
\caption{Unfolded band structures of the 21.79$^{\circ}$ twisted bilayer PtTe$_{2}$  for primitive Brillouin zone of the 0$^{\circ}$ bilayer AA stacked PtTe$_{2}$ using (a) PBE+D3, (b) vdW-optB86b and (c) r$^{2}$SCAN+rVV10 with inclusion of spin-orbit coupling. 
}
\label{fig:2179_band}
\end{figure*}

\subsection{Impact of discrepancies in DFT interlayer separations on the electronic structures}
Finally, we discuss the influence of the interlayer separations on the electronic structures for the bilayer PtTe$_{2}$  systems. Figure~\ref{fig:2179_band} displays the band structures of the twisted bilayer PtTe$_{2}$  with a twist angle of 21.79$^{\circ}$ computed with PBE+D3, vdW-optB86b and r$^{2}$SCAN+rVV10, which are unfolded to the primitive Brillouin zone of the non-twisted bilayer PtTe$_{2}$ . The band structures differ considerably, especially regarding the electronic phase: PBE+D3 yields a semimetallic band structure while the vdW-optB86b and r$^{2}$SCAN+rVV10 predict the twisted bilayer to be a semiconductor with a band gap of 0.08 eV and 0.32 eV,
respectively. 
The discrepancy in the band structures for the twisted bilayer is attributed to the huge discrepancy of the optimized interlayer separations as seen in Fig.~\ref{fig:2179_binding_curve}(d) because it governs the magnitude of the splitting between the anti-bonding and bonding states by the interlayer Te-p$_{z}$ hybridization that results in band gap reduction from the monolayer band gap. 
Indeed, due to the same mechanism, AB and AC stackings also show noticeable differences in the band structures depending on the functional, as can be seen in Fig.2 of Supplemental Material~\cite{SI}.
However, the computed band structures for the AA stacking hardly differ because of the smallest spread of the DFT interlayer separations. This suggests that the interlayer separation itself is a more sensitive factor than the choice of the DFT functionals in determining the band structures of the PtTe$_{2}$ system. Consequently, the difference in the band structure for the twisted bilayer can also be considered to originate from the discrepancy of the high-symmetric bilayer stacking modes.
This highlights again the significance of the stacking-dependent interlayer coupling to understand the twisted bilayer system, which would be particularly important for the system where the interlayer separations would lead to different electronic states such as twisted bilayer PtTe$_{2}$  system. 

\section{Conclusion}
We have investigated interlayer coupling in bilayer PtTe$_{2}$  systems by carrying out both DMC and DFT calculations, which aid in understanding the strong sensitivities this system displays with respect to commonly chosen vdW functionals. 
Our analysis serves to inform the community regarding the most reliable DFT functionals for studying the twisted bilayer systems. 
To this end, we compared the interlayer binding energies and separations between DMC and several different DFT functionals for three different high-symmetry bilayer stacking modes (AA, AB, and AC). From DMC, we find that the interlayer separations strongly depend on the stacking mode, reflecting different interaction characteristics among the stackings. 
The range of DMC-benchmarked stacking-dependent interlayer couplings is found to be well captured by only r$^{2}$SCAN+rVV10 for both energetics and interlayer separations, which makes it robust even in predicting the interlayer separations for the 21.79$^{\circ}$ twisted bilayer. This can be understood from the ability of r$^{2}$SCAN+rVV10 to correctly describe the interlayer couplings of all the local stacking modes that the twisted bilayer contains. 
Furthermore, we found differing degrees of discrepancy in the DFT interlayer separations depending on the stacking modes, with the largest discrepancies in interlayer separation and band structure appearing for the less stable AB and AC stackings. This behavior directly leads to a huge difference in the properties of the twisted bilayer.
Our results underscore the significance of understanding the interlayer couplings of high-symmetry bilayer stacking modes beyond the ground state configuration, including the less stable stackings that appear under twist. 

In this respect, our DMC calculations play a guiding role in not only narrowing down large discrepancies among vdW-DFT functionals for systems with complicated interactions in PtTe$_{2}$ but also in providing comprehensive insight into critical ingredients to model the twisted bilayer system, namely, stacking-dependent interlayer coupling. 
It is revealed from the comparison between the GGA-based and meta-GGA-based vdW-DFT results against the DMC calculations that improving the semi-local exchange-correlation potentials provides a pathway to develop more sophisticated DFT functionals that capture the stacking-dependent interlayer interactions of complicated systems.
Furthermore, r$^{2}$SCAN+rVV10, which we find to be the most reliable functional for the bilayer PtTe$_{2}$  system, shows promise to be utilized in further theoretical studies of PtTe$_{2}$ systems, such as twist-induced phase transitions which are of great current theoretical interest.



\section*{Acknowledgements}
\label{sec:acknowledgments}

Work performed by J. A., and J. T. K. (concept, DMC and DFT total energy calculations, data analysis, manuscript writing) was supported by the U.S. Department of Energy (DOE), Office of Science, Basic Energy Sciences, Materials Sciences and Engineering Division, as part of the Computational Materials Sciences Program and Center for Predictive Simulation of Functional Materials.
The work performed by S.-H. K. and M. Y. (band unfolding calculations, data analysis, manuscript writing) were supported by the U.S. DOE, Office of Science, National Quantum Information Science Research Centers, Quantum Science Center (S.-H. K.) and U. S. DOE, Office of Science, Basic Energy Sciences, Materials Sciences and Engineering Division (M. Y.). 

An award of computer time was provided by the Innovative and Novel Computational Impact on Theory and Experiment (INCITE) program. This research used resources of the Argonne Leadership Computing Facility, which is a DOE Office of Science User Facility supported under contract DE-AC02-06CH11357. This research also used resources of the Oak Ridge Leadership Computing Facility, which is a DOE Office of Science User Facility supported under Contract DE-AC05-00OR22725 and resources of the National Energy Research Scientific Computing Center, a DOE Office of Science User Facility supported by the Office of Science of the U.S. Department of Energy under contract no. DE-AC02-05CH11231 using NERSC award BES-ERCAP0028956.
 
This manuscript has been authored by UT-Battelle, LLC under Contract No. DE-AC05-00OR22725 with the U.S. Department of Energy. The United States Government retains and the publisher, by accepting the article for publication, acknowledges that the United States Government retains a non-exclusive, paid-up, irrevocable, worldwide license to publish or reproduce the published form of this manuscript, or allow others to do so, for United States Government purposes. The Department of Energy will provide public access to these results of federally sponsored research in accordance with the DOE Public Access Plan (http://energy.gov/downloads/doe-public-access-plan).

\section{Reference}
\bibliography{main_new2.bbl}

\end{document}